\documentstyle[aps,epsfig,floats]{revtex}

\begin{document}
\parskip 2mm
\draft

\twocolumn[\hsize\textwidth\columnwidth\hsize\csname@twocolumnfalse%
\endcsname

\title{Pore Stabilization in Cohesive Granular Systems}
\author{Dirk Kadau$^{1}$, Guido Bartels$^1$, Lothar Brendel$^2$, 
        and Dietrich E. Wolf$^1$\\[2mm]}


\address{$^1$ Theoretische Physik, Fakult{\"a}t 4,
     Gerhard-Mercator-Universit{\"a}t Duisburg,
     47048 Duisburg, Germany}
\address{$^2$ LMGC, UMR CNRS 5508, Universit{\'e} de Montpellier II,
     CC 048,  F-34095 Montpellier Cedex 05, France}

\date{\today}
\maketitle

\begin{abstract}
Cohesive powders tend to form porous aggregates which can be compacted
by applying an external pressure. This process is modelled 
using the Contact Dynamics method supplemented with a cohesion law and
rolling friction.
Starting with ballistic deposits of varying density, we investigate
how the 
porosity of the compacted sample depends on the cohesion strength and the
friction coefficients. This allows  to explain different pore
stabilization mechanisms. 
The final porosity depends on the cohesion force scaled by the
external pressure and on the lateral distance between branches of the
ballistic deposit $r_{\rm capt}$. Even if cohesion is switched
off, pores can be stabilized by Coulomb friction alone. This effect is
weak for round particles, as long as the friction coefficient is
smaller than 1. However, for nonspherical particles the effect is much
stronger.
\end{abstract}

]
%
%
%
%

%
%
\section{Introduction}

Cohesion plays an important role in modern powder and nano technology. 
It determines the compaction and sintering behaviour as well as the mechanical
properties like yield under shear stress \cite{morgeneyer2002}
An important tool to understand the behaviour microscopically is the
modelling on the particle scale, i.e.\ discrete element methods \cite{wolf96}. 

So far most computational studies have neglegted cohesion between
particles. This is justified in dry systems on scales where the
cohesive force is weak compared to the gravitational force on the
particle, i.e.\ for sand and coarser material, which collapses under its own
weight into a random dense packing. Powders already show cohesion
effects: With decreasing grain diameter cohesive forces lead to highly porous
media. For particle diameters in the
nanometer range the cohesive force becomes the dominant force, so that
particles stick together upon first contact.

When sintering nano-powders care has to be taken that the grains do not
grow. This is achieved in so-called sinter forging 
under high pressure at relativly low temperatures
\cite{hahn94}. In this process the highly porous powder gets compacted.
In order to simulate this compaction process we use the {\em contact
dynamics} method, which in principle \cite{unger2002} allows perfectly rigid
particles with Coulomb friction \cite{jean99,moreau94} without
regularization of these force laws. Whereas in soft particle
molecular dynamics there already exist  cohesion models
\cite{kun2001,schinner2000,roux98,luding2002}, in contact
 dynamics cohesive models are just at the beginning
\cite{radjai2001,jean2001,kadau2002}.

Here we present contact dynamics results for two dimensional systems of round
particles, which 
can be realized in experiments by aggregates of parallel cylinders
\cite{travers1986}. Cohesion as well as rolling friction were included,  
which turns out to be crucial to describe the high porosity of nano powders. 
In three dimensions, torsion friction is needed in addition \cite{bartels2002}.
Here we explain how the different contact properties contribute to stabilizing
the pores.


\section{Contact Dynamics Simulaton}
\subsection{ Without Cohesion and Rolling Friction}

It is believed that the physics of dry dense granular matter with many lasting
contacts is determined by volume exclusion and Coulomb friction. Both contact
laws are {\em nonsmooth}, i.e. they do not determine the forces as functions of
state variables (positions and velocities). The forces at a nonsliding contact
are reaction forces in the sense that they have to compensate all external
forces which would violate the constraints of volume exclusion and zero relative
velocity. The contact dynamics method \cite{jean92,moreau94,jean99} is designed
such that it determines these constraint forces in every time step. Elastic
deformations of the particles need not be taken into account, hence the choice of the
simulation time step is not coupled to the stiffness of the particles as it is
in soft particle molecular dynamics.

For two particles in contact, the constraint force can be
determined analytically, if the externally applied forces are known. But typical
dense systems consist of many 
particles involving many contacts forming \emph{contact networks}, and calculating all
contact forces is a global problem, because the force at a contact
influences the neighboring contacts and so on. To find a solution
(which is not necessarily unique, cf.\ e.g. \cite{brendel96}),
usually an iterative procedure is applied: The force at each contact is
calculated in 
random order by considering the preliminary forces of adjacent contacts
as already correct (which allows the same treatment as
for external forces). This iteration goes on until a 
convergence criterion is fulfilled. A sensible choice is to demand the relative
force changes at every contact to be below a given treshold during
a certain number of consecutive iterations.


\subsection{Cohesion Model}
%
%
\begin{figure}
  \begin{center}
    \epsfig{file=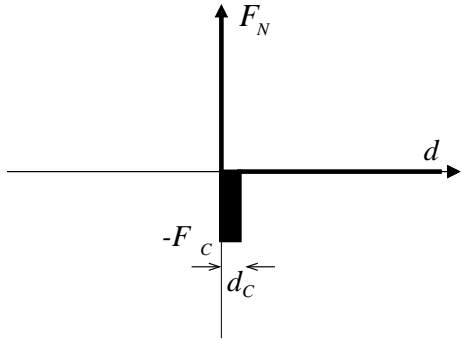, width=0.7\linewidth, angle=0}
  \end{center}
  \caption{Signorini graph (normal force $F_{\rm N}$ vs.\ distance $d$)
    for a cohesive contact: Distances above zero are allowed while
    interpenetration is avoided by means of $F_{\rm N}>0$. In contrast
    to the usual Signorini graph, a negative force
    (attractive force) is allowed within the cohesion range $d_{\rm C}$.
    (Fat lines show the allowed values.) }
  \label{fig:signorini_c_n}
\end{figure}
%
%
%
\begin{figure}
  \begin{center}
    \epsfig{file=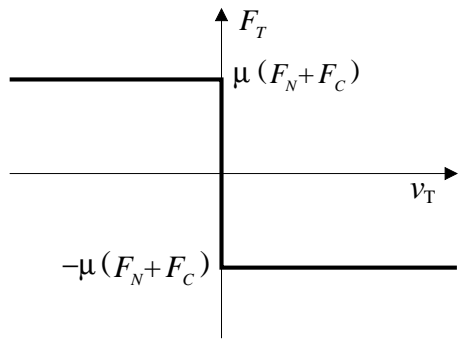, width=0.7\linewidth, angle=0}
  \end{center}
  \caption{Coulomb graph (tangential force $F_{\rm T}$ vs.\ tangential velocity
    $v_{\rm T}$) for a cohesive contact: The tangential force $F_{\rm
      T}$ hinders the contact from sliding up to the Coulomb
    threshold, above the contact starts sliding. This threshold is
    increased due to the ``normal'' cohesion. (Fat lines show the
    allowed values.) }
\label{fig:coulomb_c_n}
\end{figure}
In this section the cohesion model we use (``normal cohesion'') is described. 
For simplicity it has only two parameters, a constant attractive cohesion
force $F_{C}$ in normal direction, which acts over a short range $d_C$
determining the cohesion energy $E_C=F_C d_C$ (Fig.~\ref{fig:signorini_c_n}).
The particles are still considered as perfectly rigid: The cohesive 
interaction does not deform them. A contact can only open, if 
an external pulling force exceeds the threshold $F_C$ and performs work
$E$ larger than $E_C$, so that the particles separate with a 
kinetic energy $E-E_C$.
While Radjai {\em et al.} \cite{radjai2001} use the same cohesion
model, Jean {\em et al.} \cite{jean2001} proposed a different
implementation, the \emph{FCR model}, which has more parameters.

The introduction of a \emph{force scale} $F_C$ into
Signorini's condition (Fig.~\ref{fig:signorini_c_n})
leads to a numerical complication. 
This is related to the  existence of ``shocks'': 
Because of the perfect rigidity of the particles, a finite
momentum $\Delta p$ can be transmitted instantaneously, if the
connected cluster of particles, to which the contact belongs, 
collides with some other particle or cluster. 
This corresponds to an
infinite normal contact force in form of a Dirac-pulse $F_N=\Delta
p\,\delta(t)$. Due to the time discretization by
steps of $\Delta t$, though, one gets a finite $F_N=\Delta
p/\Delta t$ during the time-step containing the shock, which cannot  be
distinguished from a force evolving continuously at a lasting, non-shocked contact.
Whereas for $F_C=0$
shocks and persisting contacts can be treated in the same
manner \cite{jean99}, it depends on $\Delta t$, whether a contact with given
$F_C > 0 $ and $E_C > 0$ opens or not, if it is shocked ($\Delta p<0$). 
For large $\Delta t$ it may happen, that $|\Delta p / \Delta t|$ drops
below the threshold $F_C$, so that the contact cannot open. Indeed we
find that the number of contacts which open in a simulation run
decreases with increasing $\Delta t$. 

However, this is not the only source of systematic errors related to
the discrete time step: Probably most important is that the iterative
determination of the forces is only accurate within a given tolerance,
so that the constraints are not perfectly obeyed. Hence a time step
in general leads to a small overlap between the particles, which
depends on $\Delta t$. The third systematic error occurs in time steps
during which a contact opens. As the cohesion force is acting over the
whole time step, although $d=d_C$ is reached some time in between, the
cohesion energy dissipated during opening the contact is slightly
overestimated. All three systematic errors become smaller for smaller
time steps. 

\begin{figure}
  \begin{center}
    \epsfig{file=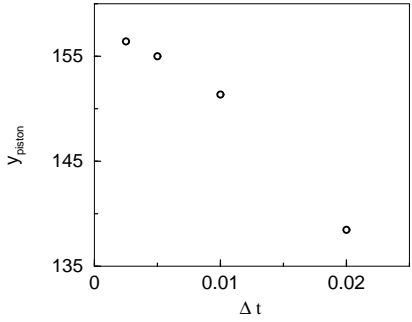, width=0.5\linewidth, angle=270}
  \end{center}
  \caption{The final position of the compacting  piston after
    consolidation is influenced by the simulation time step used.  }
\label{fig:yfinal_at_dt}
\end{figure}

As an example the final piston position as a result of a
simulation of uniaxial consolidation by a fixed external force is
shown in Fig.\ref{fig:yfinal_at_dt} as a function of the time step.
For decreasing time step the simulation result systematically
increases. In our simulations in the next sections we use a time step
$\Delta t = 0.01$, which leads to an average overlap between the
particles of about 5\% of their radius. This explains why in 
Fig.\ref{fig:yfinal_at_dt} the final piston position for $\Delta t
=0.01$ is about 5\% smaller than the extrapolated value at $Delta t =
0$. To get a lower percentage one would have to use smaller time steps
which leads to an increase of computation time. We only compare
simulation results which were obtained with the same time step.

%
%
%

We also did quasistatic simulations, where shocks are eliminated by
resetting the velocities to zero after every time step. Then the results
are different from the case of finite compaction speed:
One gets more porous structures within the compaction process by a
piston with an external load. Due to neglecting inertia effects the
system's reaction force onto the piston is below the external force on
the piston at any time. In the ``end'' both forces are almost equal,
but a low difference is left leading to a sustained movement of the
piston. Thus, the final static configuration as found in the
simulations presented in this paper, cannot be reached via a
quasistatic simulation in finite computation time. This simulation
methods would be applicable for simulating systems under constant
strain only. 

The opening of a contact needs usually several time steps, in which
the pulling force exceeds $F_C$. The
pulling force minus $F_C$ has to perform the work $E_C$ necessary to
reach the distance $d=d_C$, where the contact breaks. 
In our model a contact which started to open, but at which $d=d_C$ has
not yet been reached, is not pulled back by the cohesive force, if the
{\em pulling} force becomes weaker than $F_C$
(concerning nano-particles a sinter neck which was pulled  by an
external force to a thinner neck is not built up to its former
width again without sintering). Such a weakened but not yet broken
contact can only be strengthened again (decrease of $d$), if the particles
are {\em pushed} together externally. 

For comparison we also did some simulations with a cohesion model, in
which the cohesion force is able to pull the particles together again,
if the contact had not been broken. 
A contact which has been slightly opened by an external force (i.e.\
forces from the other contacts on the particles) is pulled back to
zero gap between the particles again (the normal force not exceeding
the threshold $-F_C$). As a result a contact, which did not break,
has a chance to relax, instead of remaining weakened. 
Simulating this model gives  similar results
as presented in this paper. 
The main difference is due to the roughness of the upper envelope of
the ballistic deposits which we choose as initial configuration for
the simulation of the compaction process (Fig.~\ref{fig:large}).
For high cohesion values the system 
really stays nearly unchanged, whereas in the simulations presented
in the following the system compacts, until the upper envelope has
become flat.

Cohesion does not only counteract the
opening of a contact but also hinders it to become sliding.
Therefore the Coulomb friction law needs to be modified.
In order to avoid introducing additional parameters, we assume
that the Coulomb friction is simply raised to $\mu (F_N+F_C)$
like in \cite{radjai2001} (Fig.~\ref{fig:coulomb_c_n}).
An alternative modification would be to increase the
friction coefficient $\mu$ depending on $F_C$. In Sec.~\ref{sec:results}
we show, that this by itself would lead to the stabilization of pores
even if ``normal cohesion'' would be absent.


\subsection{Rolling Friction Model}
%
%
\begin{figure}
\begin{center}
\epsfig{file=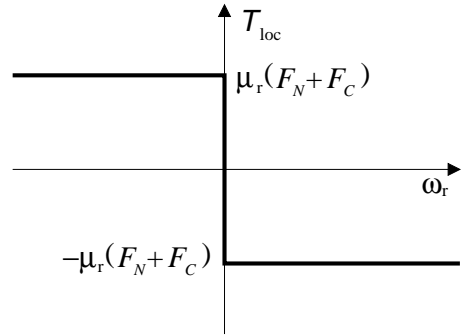, width=0.7\linewidth, angle=0}
\end{center}
\caption{Rolling condition: Similar as in the Coulomb graph the local
  torque $T_{\rm loc}$ hinders the two particles from rolling against
  each other up to a threshold. Above this threshold the contact
  becomes rolling.}
\label{fig:rollingcond}
\end{figure}
In the case of round particles, rolling of two particles on each
other must be considered. Its special importance in the context of
cohesive contacts (as opposed to purley repulsive particles) can be
explained as follows: Cohesion leads to stabilization of chains (i.e.\ 
particles with a coordination number of two), and thus allows for a
less compact structure. The latter effect, however, is possible only
if rolling is suppressed, otherwise the chains are floppy and will fold.

Introducing rolling friction to the model means to allow for a local
torque. But in principle, the \emph{point} contact formed by two
adjacent rigid discs (or spheres) cannot exert a torque. This may be
compared to the fact that the microscopic origin of Coulomb's law of
friction is not obvious for a contact with zero area.
In both cases we regard the
dimensions of the contact area and deformation zone as negligibly
small compared to the particles.

The condition for rolling is similar to Coulomb's law for sliding
(Fig.~\ref{fig:coulomb_c_n}): The contact can bear a local torque up
to the threshold $\mu_{\rm r}(F_N+F_C)$ (cf.\ 
Fig.~\ref{fig:rollingcond}). If this is exceeded, the contact becomes
a rolling one and hence exhibts a local relative angular velocity
$\omega_{\rm r}$. According to experimental results
\cite{australische_gruppe99}, the coefficient for rolling friction
$\mu_{\rm r}$ is not chosen to be velocity dependent as it would be
for a viscoelastic material \cite{poeschel98},
but is regarded as a phenomenological material constant which includes
a (microscopical) length scale.

For the simulations shown in the following, we consider only
discs of identical size and mass. Only then
the equations of motion 
for the relative angular velocity $\omega_{\rm r}$ on the one hand,
and 
for the relative translational velocity on the other, are not coupled,
which simplifies the implementation of rolling friction significantly.


\section{Simulations Results} \label{sec:results}
\subsection{Initial Configuration}
In principle there are two general methods to produce nano-particles:
the first is to fracture bulk material. The second one builds
nano-particles by chemical precipitation or condensation in a liquid
or gas environment. While the production in large quantities is
usually done in liquids, the production in the gas phase has some
advantages (higher pureness of the material and sharper grain size
distribution). 
%
%
%
\begin{figure}
\begin{center}
\epsfig{file=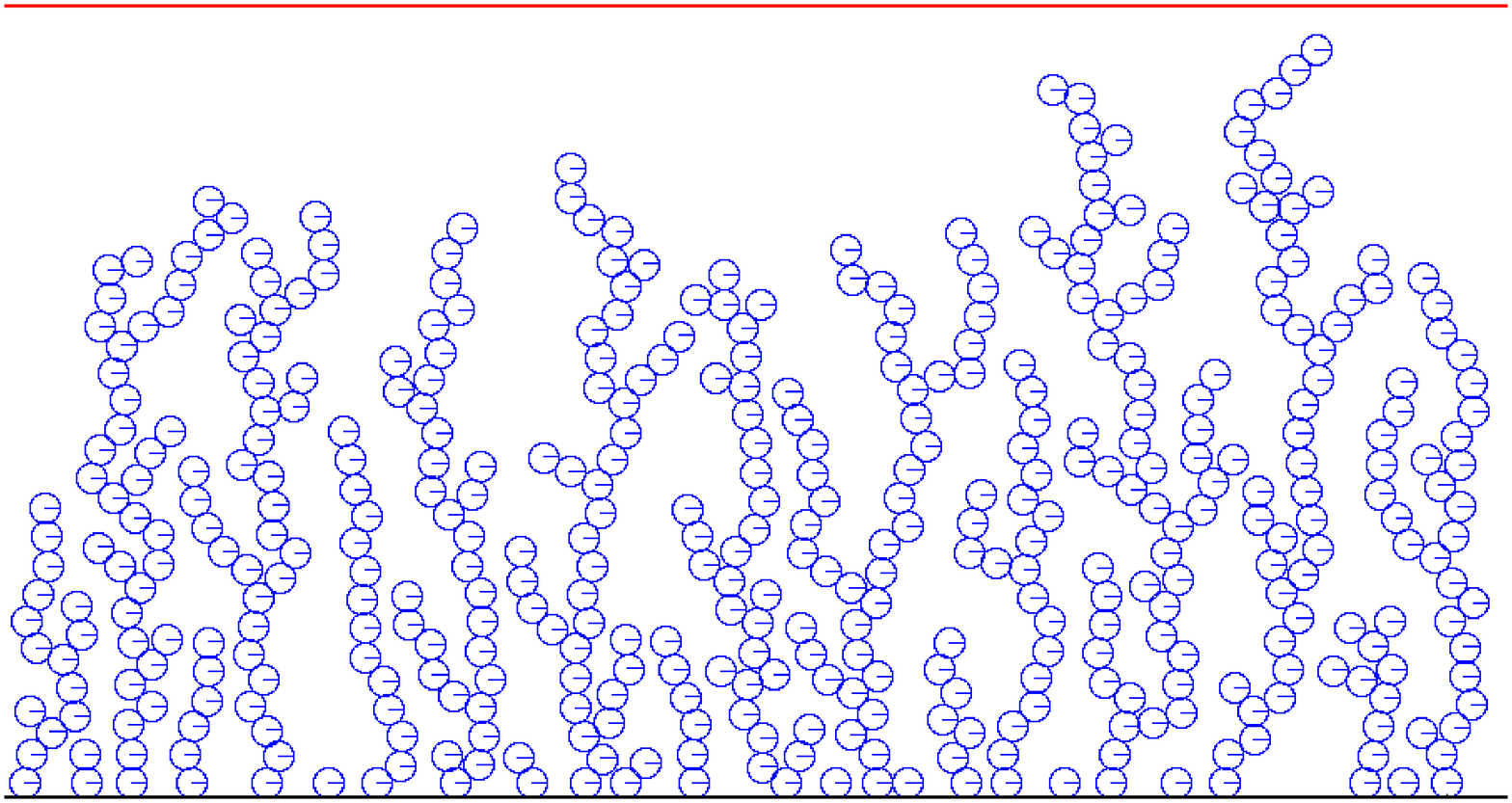, width=\linewidth, angle=0}
\end{center}
\caption{Initial configuration: identical sperical particles are
  ballistically deposited.}
\label{fig:small_ini}
\end{figure}
To extract
nanoparticles out of a gas flow one can use a filter where the
particles are deposited on. Simulations of the filter process by a
fiber network show a finger-like structure of the deposit
\cite{lantermann,filippova97}. Similar structures are
obtained on a flat surface by ballistic deposition: Particles fall
from the top randomly and stick to the first particle they reach
within a capture radius $r_{\rm capt}$ around the particle. Such a
ballistic deposit is 
shown in figure~\ref{fig:small_ini} for two dimensions \cite{meakin91}.
These structures are not fractal, i.e. the porosity is not depending
on the system size but on the capture radius. 
%
%
%
\begin{figure}
\begin{center}
\epsfig{file=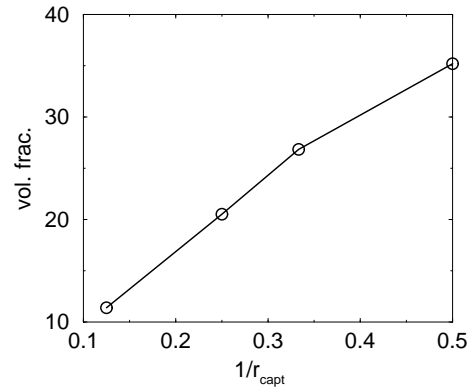, width=0.6\linewidth, angle=270}
\end{center}
\caption{ The density (volume fraction) of the ballistic deposited configurations
  decreases with increasing capture radius (here plotted against
  $1/r_{\rm capt}$). } 
\label{fig:dens_at_rcapt}
\end{figure}
The capture radius is a measure for the average distance between the
 branches so that the volume fraction of a ballistic deposit
 increases about linearly with the reciprocal of the capture radius
 (Fig.~\ref{fig:dens_at_rcapt}). For
monodisperse spherical particles the minimum capture radius is two
particle radii, i.e.\ the minimal distance between the centers of mass
two particles can have. 
 In the following sections different ballistic deposits with
periodic boundary conditions in lateral direction are compacted by a
piston with constant load in vertical direction.

\subsection{Influence of different contact properties}\label{sec:contact_props}
%
%
\begin{figure}
\begin{center}
\epsfig{file=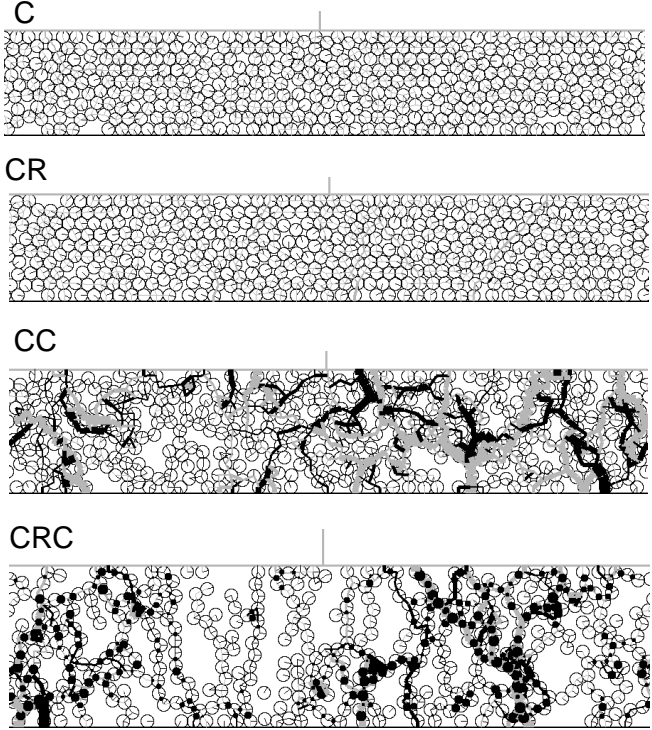, width=\linewidth, angle=0}
\end{center}
\caption{Configurations after compaction with different
  contact properties: Coulomb friction only (C);
 Coulomb and rolling friction (CR);
 Coulomb friction and cohesion (CC);
 Coulomb and rolling friction and cohesion (CRC); gray and black lines
 show compressive and tensile forces, respectively. The line thickness
 is a measure for the force value. The size of the black dots
 at the contacts indicates the magnitude of the opposite torques exerted
 by the contact on the adjacent particles.
  }
\label{fig:small_fin}
\end{figure}
Initial configuration is a ballistic deposit with capture radius of
$2.5$ particle radii (Fig.~\ref{fig:small_ini}) containing $433$
particles. After the compaction process by applying an external force
on the upper piston one finally reaches a final equilibrium
state. Figure~\ref{fig:small_fin} shows the final state for different
contact properties used within the simulation. If one uses Coulomb
friction (C) only (in addition to excluded volume interaction) one
ends up with a high volume fraction (78\%). Using rolling friction in
addition (CR) a volume fraction of 77\% is reached. Compared to a
perfect triangular lattice with volume fraction $\pi /\sqrt{12}\approx
90.7\%$ there are only a few larger pores in these two
configurations. This can be explained by the fact that force chains
must be stabilized by adjacent particles avoiding the lateral movement
of the chains \cite{radjai98}. Coulomb friction and
cohesion (CC) lead to larger pores (see Fig.~\ref{fig:small_fin}) and
thus a volume fraction of 68\%. The stabilization mechanism of force
chains (and thus pore stabilization) is a different one: Force chains
of compressive forces (gray lines) are stabilized by nearby
force chains of tensile forces (black lines) so that the piston load
is carried by the system. The higher thickness of the force lines
indicate that the stresses in the system are much higher than in the
absence of cohesion. However, unhindered rolling leads to a
destabilization of single particle chains. Therefore additional
rolling friction (CRC) leads to the highest porosity (volume fraction
of 51\%) because the system includes stable single particle chains.
In this case there is
stabilization of three degrees of freedom: the separation of the
particles, the lateral movement as well as the rolling against each
other. These three degrees of freedom are stabilized 
by the combination of cohesion, Coulomb friction, and
rolling friction.


\subsection{Compaction process}
%
%
%
\begin{figure}
\begin{center}
\epsfig{file=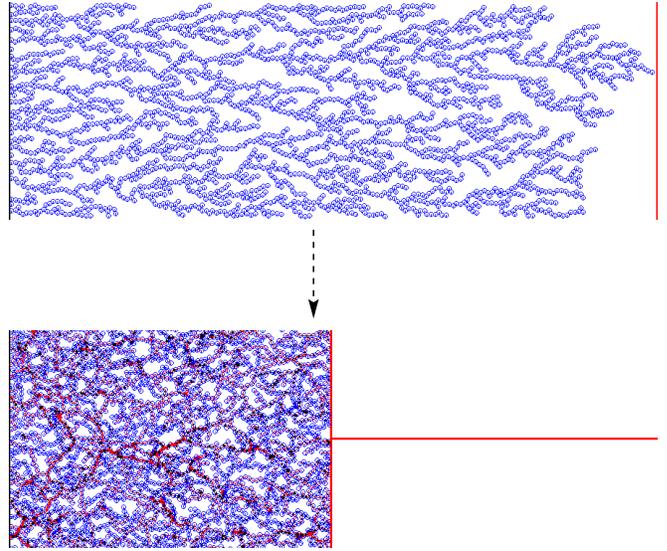, width=\linewidth, angle=0}
\end{center}
\caption{Ballistic deposit before and after compaction by
  a constant external force on the  piston. Contact properties
  include Coulomb friction, rolling friction and cohesion.
  }
\label{fig:large}
\end{figure}
%
%
%
%
\begin{figure}
\begin{center}
\epsfig{file=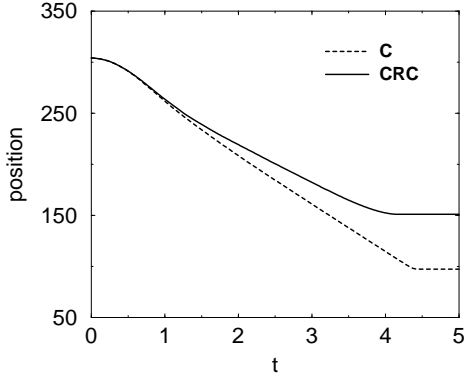, width=0.6\linewidth, angle=270}
\end{center}
\caption{Time evolution of the piston's position during the compaction
 process for Coulomb friction only (C, dashed) and Coulomb and
 rolling friction and cohesion (CRC, full line) in comparison. 
  }
\label{fig:time_evolution}
\end{figure}
In order to get insight into the compaction process itself the temporal
behaviour of the system is studied in this section. We simulated a
ballistic deposit with capture radius $r_{\rm capt}=2.5$
containing $2746$ monodisperse spherical particles
(Fig.~\ref{fig:large}) with periodic boundary conditions in lateral
direction. We compared the compaction dynamics for two different
contact properties between the particles: coulomb friction (C) only leads
to the most compact final configuration, whereas Columb, rolling
friction and cohesion (CRC) lead to the most porous final
configuration (see sec.~\ref{sec:contact_props}). In both cases the
time evolution of the piston position
(Fig.~\ref{fig:time_evolution}) shows three phases of the compaction
process. In the first phase the piston accelerates. Initially it is in
contact with only one branch of the ballistic deposit. Only later it 
sweeps up substantial mass. Then, in the second phase 
the piston moves with a nearly constant velocity. The momentum
transfered to the system by the force acting on the piston (in the
cohesive case partly compensated by a reaction force of the opposite
side of the container) is consumed by a 
linearly increasing mass swept up by the piston:
\begin{equation}
F_{\rm piston} - F_{\rm reaction} = \dot{p} = \dot{m}\cdot v
\end{equation}
In the third phase the
piston finally reaches a constant position (only neglectable small
oscilations due to quasi-elasticity \cite{unger2002}): 
$F_{\rm piston} = F_{\rm reaction}$. The final position of the piston
is different due to the different stabilization mechanism
(see sec.\ref{sec:contact_props}). 

\subsection{Influence of cohesion strength}
%
%
%
\begin{figure}
\begin{center}
\epsfig{file=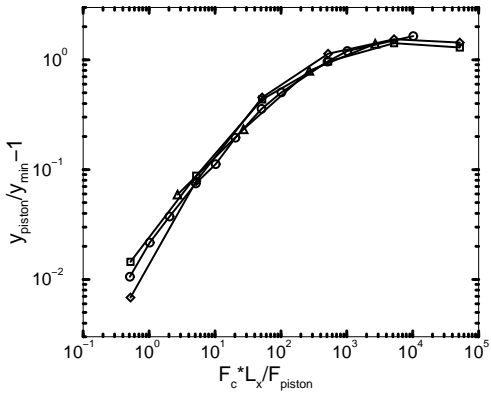, width=0.6\linewidth, angle=270}
\end{center}
\caption{Scaling plot for different cohesion strength shows alignment
 for different Systems. The systems are different in size and aspect
 ratio but have the same capture radius for ballistic deposition and
 thus approximately the same initial density. 
 }
\label{fig:scaling}
\end{figure}
The final piston position of different systems compactified by a constant
external force is analyzed. At the contacts 
Coulomb and rolling friction, as well as cohesion with
different values of $F_C$ are acting. The
initial systems are ballistic deposits with capture radius $r_{\rm
 capt}=2.5$ and periodic boundary conditions. They consist of
different numbers of particles and have
different aspect ratios. We found that all the data collapse onto a
single curve, Figure~\ref{fig:scaling},
if the final piston position is scaled by  the final
piston position without cohesion, $y_{\rm min}$, and 
the cohesive force $F_{\rm C}$ by the pressure exerted by the piston,
$F_{\rm piston}/L_x$ in the two dimensional case.
In the beginning
an (almost linear) increase can be seen. For large cohesive force the
scaled final piston position reaches a constant value.
In this region there is almost no compaction. 
The fact that $F_{\rm C} L_x/F_{\rm piston}$ is the relevant quantity
for the compaction process
implies that the typical distance of strong force lines is not
depending on the system size, as is also found for systems without cohesion
\cite{radjai98}.

\subsection{Scaling behaviour for different initial densities}
%
%
\begin{figure}
\begin{center}
\epsfig{file=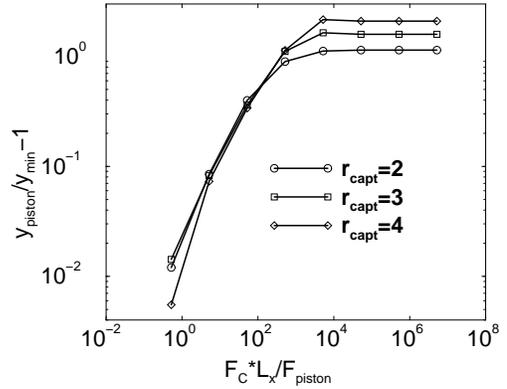, width=0.6\linewidth, angle=270}
\end{center}
\caption{Scaling plot for different cohesion strength shows alignment
 for Systems with different capture radius for small cohesion pressure
 ratios only (cp.\ Fig.~\ref{fig:scaling}). For strong cohesion the
 final densities are mainly given by the initial densities.
 }
\label{fig:scaling_rcapt}
\end{figure}
In the previous section the compaction of 
ballistic deposits with the same capture radius was
investigated. Now three ballistic deposits of the same size
in vertical and lateral directions, but different capture radius are
compacted by a piston with a fixed external load. Thus the initial systems
contain different numbers of spherical particles, namely $2777$
($r_{\rm capt}=2$), $2111$ ($r_{\rm capt}=3$) and $1624$ particles
($r_{\rm capt}=4$). The initial densities  scale with $1/r_{\rm
 capt}$ (see fig.~\ref{fig:dens_at_rcapt}). The same plot as in
figure \ref{fig:scaling} (each curve averaged over 10 runs) 
leads to a data collapse for small ratios of cohesion and external
pressure on the piston only. In this region the different initial
densities do not play an important role. For high cohesion values 
the final densities remain essentially the initial densities, so that there
is no scaling in this region.

\subsection{Pore stabilization without cohesion}  
%
%
\begin{figure}
\begin{center}
\epsfig{file=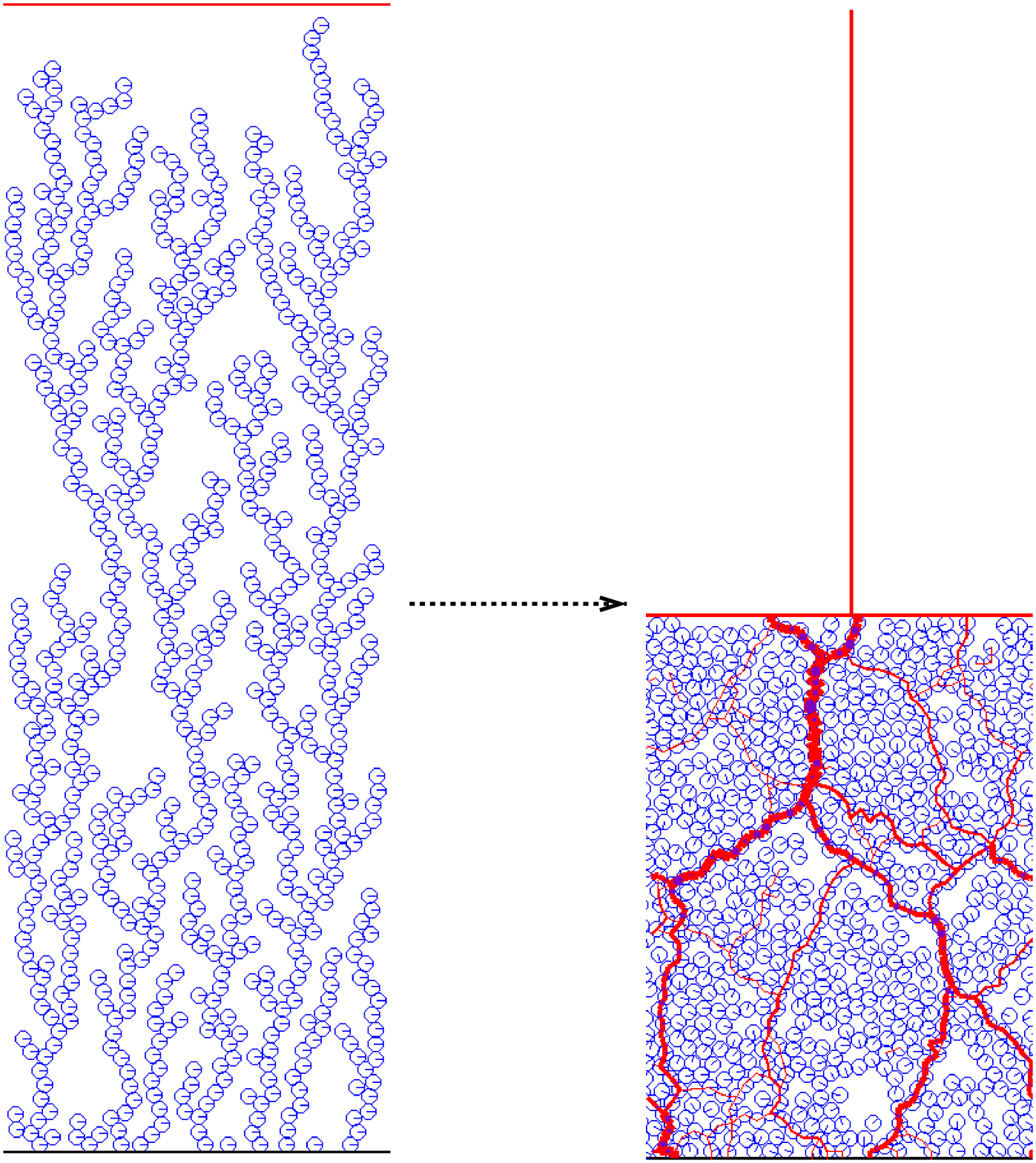, angle=270, width=\linewidth}
\end{center}
\caption{ Initial and final configuration after compaction with
 friction coefficient $\mu=3.8$ and rolling friction coefficient
 $\mu_{\rm r}=2.0$ instead of cohesion.
 }
\label{fig:medium}
\end{figure}
%
%
%
\begin{figure}
\begin{center}
\epsfig{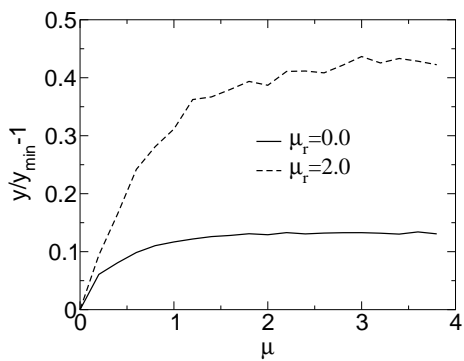}
\end{center}
\caption{ Increased friction coefficient $\mu$ leads to pore
 stabilization and thus to lower density. 
 Additional rolling friction amplifies this effect.
 }
\label{fig:density_at_mu}
\end{figure}
The results presented in Figs.\ \ref{fig:scaling} and
\ref{fig:scaling_rcapt} suggest the following physical picture: The
external load of the piston must be carried by a set of strong force
lines, which have a typical distance.  The external load per force
line is proportional to the pressure $F_{\rm piston}/L_{\rm x}$. It can
be viewed as the destabilizing force along a force line and must be
balanced by the stabilizing influence of another force, which in the
previous sections was related to cohesion (and the fixed rolling
friction), so that the scaling variable was $L_{\rm x} F_{\rm
  c}/F_{\rm piston}$. This variable had to be big enough to prevent
compaction.

The question we want to address in this section is, whether friction
forces alone can provide the stabilization as well. It is plausible to
identify the external load per force line with the typical normal
force at a contact along the force line, $F_{\rm n} \approx F_{\rm
  piston}/L_{\rm x}$. The ratio between the stabilizing friction force
$F_{\rm t} = \mu F_{\rm n}$ and the destabilizing force would then be
$L_{\rm x} F_{\rm t}/F_{\rm piston} \approx \mu$. This argument suggests
that strong enough Coulomb friction may stabilize pores. This is
indeed the case, as Fig.\ \ref{fig:density_at_mu} shows. This effect
is not very high (maximally about 10\%), because the Coulomb friction
cannot provide stabilization against buckling of the force lines.
Therefore strong force lines need weak forces from the side as pointed
out in \cite{radjaiPRL98}.  Alternatively, rolling friction may
stabilize the force lines against buckling. In combination with
Coulomb friction this allows much larger pores, as the dashed curve in
Fig.\ \ref{fig:density_at_mu} shows. However, the pore geometry is
totally different from the one obtained in cohesive materials (compare
Figs.\ \ref{fig:large} and \ref{fig:medium}). In the absence of cohesion
large pores are found underneath arches. Presumably strong force lines
stop the motion of grains on the upper side, while below pores open up
due to inertia effects. Cohesion would lead to correlated motion of
clusters of grains and would also prevent the inertial rupture of the
structure underneath an arch.

%
%
\begin{figure}
\begin{center}
\epsfig{file=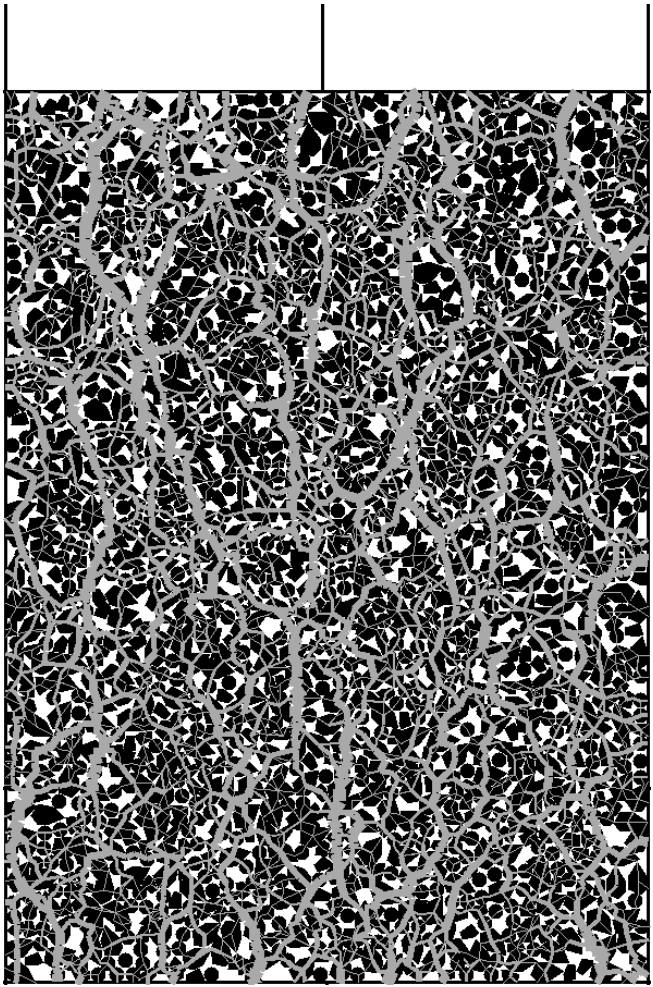, angle=0, width=\linewidth}
\end{center}
\caption{ Final configuration after compaction with
 friction coefficient $\mu=0.5$  for a mixed system of convex polygons
 and discs. 
 }
\label{fig:mixed_coul}
\end{figure}
%
%
\begin{figure}
\begin{center}
\epsfig{file=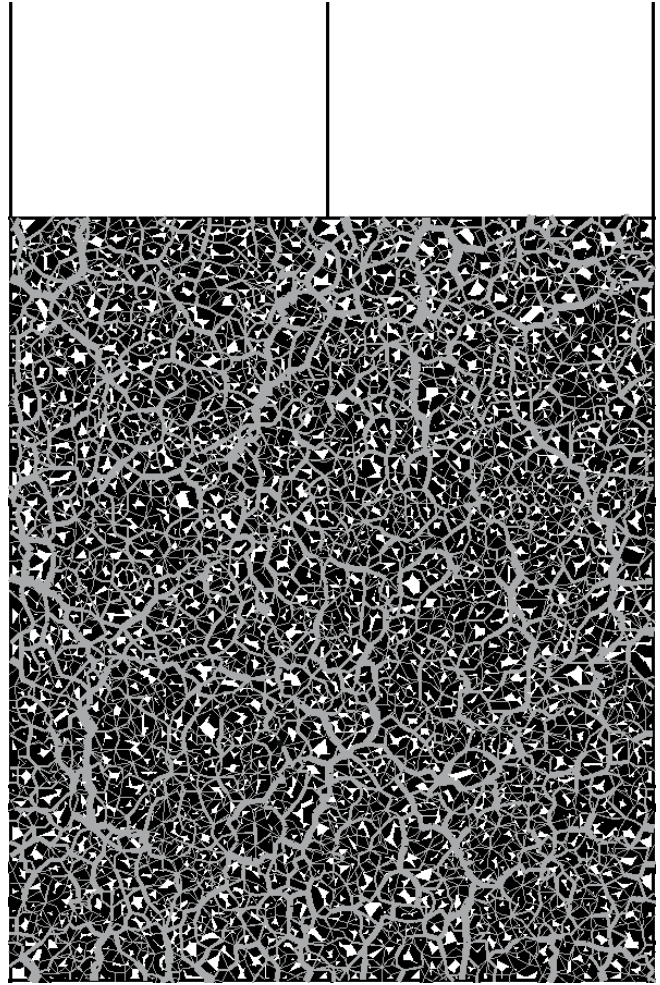, angle=0, width=\linewidth}
\end{center}
\caption{ Final configuration after compaction without Coulomb
  friction 
 for a mixed system of convex polygons and discs. The system ends in a
 less porous structure as after compaction with Coulomb
 friction.
 }
\label{fig:mixed_no_coul}
\end{figure}

Without the use of rolling friction higher porosity will be reached by
the use of non-sperical particles. To show this effect we simulated a
a system of about $1700$ particles consisting of two different
 convex polygonal particle types and one spherical particle type. The
 diameter of the particles is  chosen within the same range. The
 initial configuration is a random loose packing, where no particles
 are in contact. After precompaction to a denser state the system is
 compacted by a piston with constant load using Coulomb friction
coefficient $\mu =0.5$ (Fig.~\ref{fig:mixed_coul}), respectively without
Coulomb friction (Fig.~\ref{fig:mixed_no_coul}). The  two
final
 configurations (Fig.~\ref{fig:mixed_coul},~\ref{fig:mixed_no_coul})
have different porosity. It is higher, if Coulomb friction is
present.
 Due to the shape disorder this effect is stronger than for the system
 of spherical particles. Here one ends up with a value $y/y_{\rm
   min}-1\approx 0.17$ compared to the value $y/y_{\rm min}-1\approx
 0.09$ at $\mu=0.5$ for the ballistic deposits of monodisperse sperical
 particles (Fig.~\ref{fig:medium}). One concludes that the corners of
 the particle have a similar effect as rolling friction for spherical
 particles. 

Thus, for cohesive non-sperical grains one expects a higher porosity
depending on the shape of the particles, due to a  similar
stabilization mechanism as we found for sperical particles using
rolling friction.


\section{Summary.}
Different contact properties lead to different configurations after
compaction. Using only Coulomb and rolling friction the final packing
is an almost compact structure. The effect of cohesion on the
porosity is higher if one uses Coulomb and rolling friction in
addition. In that case single particle chains are stable and thus the 
packing is highly porous. Important is the strength of the cohesion:
Low cohesive forces lead to similar packings as without
cohesion. Important is the relation between the stabilizing (cohesion) and
destabilizing (pushing force) forces. Here one has to take into
account that forces are extremly inhomogeneously distributed in the
system so that one must devide the external pushing force by the
average distance between strong force lines. This picture is confirmed
by applying it to a system with increased values for the friction
coefficient (above $1$), which also leads to higher porosity.


\section{Acknowledgements}
Useful conversations with T.\ Unger,  M.\ Morgeneyer, J.\ Kert{\'e}sz
J.\ Schwedes, Z.\ Fark{\'a}s and
H.\ Hinrichsen are gratefully acknowledged.
We acknowledge support by DFG within SFB 445 {\em Nano-Particles
 from the Gas-Phase: Formation, Structure, Properties} and
 within the DFG grant WO 577/3-1 {\em Compaction and Mechanical
   Properties of cohesive bulk solids}.


\end{document}